\documentclass[11pt,english,aps,manuscript]{article}
\usepackage[latin9]{inputenc}
\usepackage{geometry}
\usepackage{amsmath}
\usepackage{amssymb}
\usepackage[numbers]{natbib}
\usepackage{babel}
\usepackage{xcolor}

\newcommand{\roughly}[1]{\mathrel{\raise.3ex\hbox{$#1$\kern-0.85em
\lower1ex\hbox{$\sim$}}}}

\newcommand{\lsim}{\roughly<}

\def\cL{{\cal L}}
\def\cM{{\cal M}}
\def\cO{{\cal O}}

\def\cV{{\cal V}}

\def\mfK{{\mathfrak{K}}}

\newbox\charbox
\newbox\slabox
\def\slsh#1{{      % Feynman slash
        \setbox\charbox=\hbox{$#1$}
        \setbox\slabox=\hbox{$/$}
        \dimen\charbox=\ht\slabox
        \advance\dimen\charbox by -\dp\slabox
        \advance\dimen\charbox by -\ht\charbox
        \advance\dimen\charbox by \dp\charbox
        \divide\dimen\charbox by 2
        \raise-\dimen\charbox\hbox to \wd\charbox{\hss/\hss}
        \llap{$#1$}
}}

\def\exd{{\hbox{d}}}

\def\cL{{\cal L}}
\def\cM{{\cal M}}
\def\cO{{\cal O}}

\def\bea{\begin{eqnarray}}
\def\eea{\end{eqnarray}}
\def\be{\begin{equation}}
\def\ee{\end{equation}}

\def\ssI{{\scriptscriptstyle I}}

\def\ssN{{\scriptscriptstyle N}}

\def\ssR{{\scriptscriptstyle R}}

\def\KK{{\scriptscriptstyle KK}}

\def\CMB{{\rm CMB}}

\def\({\left(}
\def\){\right)}

\def\pref#1{(\ref{#1})}

\begin{document}

`\vspace{1cm}
\begin{center}
\textbf{\LARGE{}Cosmological Trans-Planckian Conjectures\\ are not Effective}{\Large\par}
\par\end{center}

\begin{center}
\vspace{0.6cm}
 % \vskip 1.2cm
 
 { \bf C.P.~Burgess,$^{a,b}$ S. P. de Alwis$^c$ and F. Quevedo$^d$}

 \vskip 0.6cm
 
 {\small {\it $^{a}$Department of Physics $\&$ Astronomy, McMaster University\\
1280 Main Street West, Hamilton Ontario, Canada L8S 4M1}} \\ %\\ and \\
%\vspace{.15cm}
  {\small {\it $^{b}$Perimeter Institute for Theoretical Physics \\
31 Caroline Street North, Waterloo Ontario, Canada N2L 2Y5}}\\
 {\small {\it $^{c}$Physics Department, University of Colorado, Boulder, CO 80309 USA}}\\
 %\vspace{.15cm}
{\small {\it $^{d}$DAMTP, Centre for Mathematical Sciences, Wilberforce Road,  Cambridge, CB3 0WA, UK}}	
 
 \vskip 1.5cm

\par\end{center}

\begin{center}
\vspace{0.3cm}
  
\par\end{center}

\begin{center}
\textbf{\small ABSTRACT} 
\par\end{center}

{%\small 
It is remarkable that the primordial fluctuations as revealed by the CMB coincide with what quantum fluctuations would look like if they were stretched across the sky by accelerated cosmic expansion. It has been observed that this same stretching also brings very small -- even trans-Planckian -- length scales up to observable sizes if extrapolated far enough into the past. This potentially jeopardizes later descriptions of late-time cosmology by introducing uncontrolled trans-Planckian theoretical errors into all calculations. Recent speculations, such as the Trans-Planckian Censorship Conjecture (TCC), have been developed to avoid this problem.  We revisit old arguments why the consistency of (and control over) the Effective Field Theory (EFT) governing late-time cosmology is not necessarily threatened by the descent of modes due to universal expansion, even if EFT methods may break down at much earlier times. Failure of EFT methods only poses a problem if late-time predictions rely on non-adiabatic behaviour at these early times (such as is often true for bouncing cosmologies, for example). We illustrate our arguments using simple {\it non-gravitational} examples such as slowly rolling scalar fields and the spacing between Landau levels for charged particles in slowly varying magnetic fields, for which similar issues arise and are easier to understand. We comment on issues associated with UV completions. Our arguments need not invalidate speculative ideas like the TCC but suggest they are not required by the present evidence.}

\newpage

\tableofcontents

%\newpage

\section{Introduction}
Everybody seems to have strong opinions about what goes on at trans-Planckian energies these days. This is harmless to the extent that these opinions make no difference to testable low-energy predictions, and the general principle of decoupling usually ensures that this is true. The decoupling principle formalizes the observation that very short-distance physics seems not to matter much when understanding long-distance physics, and helps understand why people could historically understand atoms without also understanding atomic nuclei in detail.\footnote{More precisely, decoupling states \cite{decoupling} that there exists a {\it choice} for low-energy (relevant and marginal) couplings for which states with energy $M \gg E$ only contribute to observables at energy $E$ suppressed by positive powers of $E/M$.} It seems to be an important feature of Nature and provides part of the reason science can progress at all. 

That is why seeking violations of decoupling is interesting, with success having both good and bad implications. The bad news would be that unknown physics at short scales might overwhelm (and so ruin) low-energy predictions, but the good news would be that otherwise unobservable physics might unexpectedly come within observational reach. 

Effective Field Theories (EFTs) are the tools designed to understand and exploit decoupling for physical systems that involve several very different energy scales \cite{Weinberg:1978kz,TheBook,PetrovBlechman}, and these tools have been widely tested across almost all areas of physics. They do so essentially by performing a Taylor expansion in powers of the hierarchy of scales, $E/M$, and doing so as early as possible so as to extract the most utility from the resulting simplicity. Searches for decoupling violation are therefore often cast as unexpected breakdowns of EFT methods. The key word here is `unexpected' because EFTs are always expected to break down when the hierarchy of scales involved is not large. By their very nature, EFTs always come together with a maximum energy scale above which they cease to be valid, defined as the scale where the underlying expansion in powers of $E/M$ is no longer sufficiently accurate

We know of no clear examples of decoupling failure in a controlled approximation, though controversial claims for its existence are sometimes made, often in cosmology. They arise in cosmology for several reasons. First, gravity plays a central role and the systematic development of EFT methods for quantum effects in gravity is relatively new \cite{Donoghue:1994dn,Goldberger:2004jt,Goldberger:2005cd,Cheung:2007st,Baumann:2010tm,Carrasco:2012cv,Burgess:2014eoa,Burgess:2015ajz} (for reviews see \cite{TheBook,Burgess:2003jk,Donoghue:2017ovt,Burgess:2017ytm}). A common attitude towards EFTs in cosmology is that they are an optional part of a theorist's menu, whose utility is limited to situations involving several hierarchies of scale. This would be a reasonable point of view for quantum electrodynamics (QED) and other renormalizable theories, but ignoring EFT issues is never possible for nonrenormalizable theories like general relativity (GR). EFT methods provide the only known way to quantify the size of quantum effects for such theories, and do so by showing that they arise as an expansion in powers of derivatives of fields times the underlying length scale of the nonrenormalizable coupling \cite{Weinberg:1978kz,Burgess:2003jk}. 

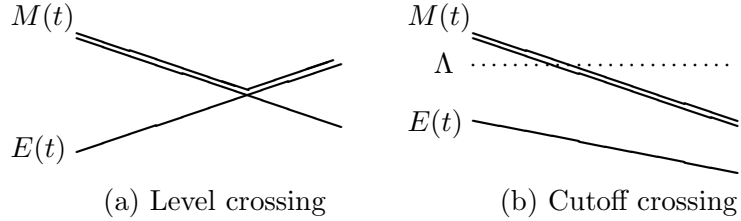
\begin{figure}
\centerline{
\begin{picture}(260,100)
\put(0,0){\begin{picture}(100,100)
    \thicklines
    \put(10,32){\line(3,1){100}}
    \put(10,75){\line(3,-1){100}}
    \put(75,56){\line(3,1){32}}
    \put(10,77){\line(3,-1){65}}
    \put(-15,80){$M(t)$}
    \put(-15,30){$E(t)$}
    \put(20,10){(a) Level crossing}
    \end{picture}}
\put(150,0){\begin{picture}(100,100)
    \thicklines
    \put(10,44){\line(5,-1){100}}
    \put(10,75){\line(3,-1){100}}
    \put(10,77){\line(3,-1){100}}
   \multiput(10,65)(4,0){25}{\circle*{1}}
    \put(-15,80){$M(t)$}
    \put(-5,62){$\Lambda$}
    \put(-15,40){$E(t)$}
    \put(20,10){(b) Cutoff crossing}
    \end{picture}}
\end{picture}
}
\caption{Adiabatic time evolution for the energy, $E(t)$ (single solid line), of an initially low-energy state, as well as the energy, $M(t)$ (solid double line), for the lightest physical UV state (with time flowing to the right). The left panel illustrates how an expansion in $E/M$ could start off being justified before eventually breaking down so EFT methods fail. The right panel describes the trans-Planckian situation where states evolve below an arbitrary but fixed cutoff, $\Lambda$ (dotted line), without UV and IR levels becoming close to one another (and so for which EFT methods need not fail). \label{fig:levcross}}
\end{figure}

The second feature about cosmology that inspires skepticism about EFT methods is the ubiquity of time-dependent background fields, since this introduces new complications whose handling within EFTs is relatively poorly developed compared with static environments. These in particular introduce new `expected' ways to leave the domain of validity of EFTs, including the occurrence of nonadiabatic evolution \cite{Burgess:2014lwa} or the failure over time of the other EFT criteria, such as $E(t)/M(t)$ becoming large even if it were initially small (see the left panel of Fig.~\ref{fig:levcross}). For surveys of these issues see \cite{TheBook,Burgess:2003jk,Burgess:2017ytm}.

The right-hand panel of Fig.~\ref{fig:levcross} illustrates a putative failure mode for decoupling during cosmology -- called the Trans-Planckian Problem (TPP) -- that has been widely discussed \cite{Martin:2000xs}. According to this picture momentum states redshift due to the universe's expansion, causing a continuous flow of states below any particular fixed UV cutoff scale (denoted $\Lambda$ in the figure). The assertion is that during a primordial inflationary period physical states that descend from above this cutoff (and so are presumably not understood) can become stretched out to become large enough to be relevant to predictions for the CMB, and this presents a problem/opportunity for calculations that are compared to CMB observations. 

The most recent iteration of these ideas regards this prospect to be sufficiently disasterous to justify a trans-Planckian censorship conjecture (TCC) \cite{Bedroya:2019snp, Bedroya:2019tba}, that states that the laws of physics must be chosen to ensure that such mode transference from UV to IR does not happen. For inflationary scenarios this is interpreted to mean that acceptable models do not permit trans-Planckian modes to become stretched out longer than the Hubble scale ($H^{-1})$, leading to an upper bound on the energy scale and duration of inflation. It provides particularly strong constraints on models with eternal inflation,\footnote{Recently \cite{Bedroya:2020rac} has argued that eternal inflation is ``marginally ruled out''  by the TCC.} for which longer periods of inflation would be more probable than shorter periods, even for a low Hubble scale during inflation.

The TCC is closely related -- and similar in spirit -- to the swampland programme \cite{Vafa:2005ui, Palti:2019pca}.  In its broadest form the swampland program asks whether robust conclusions can be drawn about the low-energy EFTs that arise from specific types of UV completions that might arise at high energies to unitarize general relativity. In practice the UV completion is usually chosen to be string theory, for want of alternatives with which sufficiently precise predictions can be made. The assertion is that not all low-energy EFTs actually have sensible UV completions, and this has led to a series of conjectures about how to identify those that do and those that don't.

Swampland conjectures so far range between those that are very likely to be true but not that informative to those that are very constraining but for which the supporting evidence is quite weak. Examples of the first type include the weak gravity conjecture \cite{ArkaniHamed:2006dz} or the proscription against the existence of global symmetries \cite{NoGlobal} (see {\it e.g.}~\cite{Burgess:2008ri} for some reasons why the prediction of no global symmetries is in the end less informative than might have been thought).  Examples of the second type include the de Sitter conjectures \cite{Obied:2018sgi}, whose main motivation is the contrast between the difficulty of finding de Sitter solutions within string theory and the relative ease with which they arise within 4D EFT candidates. (An alternative interpretation of this contrast  \cite{Burgess:2020qsc} sees it as a consequence of the generic presence of accidental scale invariances for EFTs produced by perturbative string vacua rather than pointing to a swampland. In this understanding it is also difficult to obtain de Sitter solutions within the EFTs\footnote{It is worth recalling in this context that it has been known since the 1980's \cite{Dine:1985he} that the scalar potential for moduli at arbitrarily weak string coupling $g_s$ and at arbitrarily large extra-dimensional volume $\mathcal{V}$ generically has a runaway form, and much of the progress in flux compactifications -- such as those in \cite{Kachru:2003aw, Balasubramanian:2005zx, Conlon:2005ki} --  was to allow the existence of vacua at weak enough string coupling and large enough volume to trust the approximations. But the accidental scaling symmetries ensure that the runaway form always re-emerges for arbitrarily small $g_s$ and $1/\mathcal{V}$.} once these symmetries are properly incorporated. The symmetry connection both provides intuition about the origins of various no-go theorems \cite{no-go} and suggests how these theorems can be evaded.)

Swampland conjectures have been broadly helpful in directing attention to what is generic and what is not in the low-energy EFTs arising in string theory.\footnote{Some of the arguments against de Sitter constructions (see {\it e.g.}~\cite{Sethi:2017phn}) start from the assertion that EFT methods can only be deployed for small excursions about static vacua in the UV theory, and are inappropriate for time-dependent systems (or that particular types of corrections cannot be handled when backgrounds are time-dependent). We see no evidence for this proposition -- and considerable evidence against it -- for EFTs applied to time-dependent fields in everyday non-gravitating systems \cite{Polchinski:1995ta,Burgess:2003jk,TheBook} as we discuss later.} The de Sitter conjectures have in particular focussed useful attention on the weakest points of explicit 4D de Sitter string constructions, many of which are quite complicated and have a somewhat baroque feel. These complications largely arise, however, due to the care spent maintaining control over the approximations used (such as string perturbation theory and a low-energy EFT expansion in higher dimensions), and doing so is important because the strongest arguments against the existing constructions focus on the existence of regions within the extra dimensions where some of the extra-dimensional fields leave the domain of validity of these approximations \cite{Bena:2009xk, Michel:2014lva, Polchinski:2015bea,Sethi:2017phn,Danielsson:2018ztv}. Often this occurs in strongly warped regions near the brane sources used in their construction\footnote{Some recent work \cite{Gao:2020xqh} shows that the singular regions in some of the constructions (such as in KKLT type models) need not be localized within stringy distances to the branes.}. 

Although such regions indeed complicate a full demonstration of stringy provenance, such as by taking the full description of these specific regions outside of the domain of validity of the 10D EFT, it need not invalidate the conclusions drawn from the rest of the solution unless these strongly depend on the details of these singular regions. After all, one does not need to discard all we know about atomic energy levels just because the Coulomb field of the nucleus diverges when extrapolated to the nuclear centre. We know the effective description breaks down there, but does so in a specific region, and this only has small controlled effects on atomic energy levels. In our opinion the evidence in favour of a de Sitter conjecture remains weak, but we believe it has played a constructive role by triggering numerous efforts to re-examine and improve de Sitter constructions in string theory, in both the so-called KKLT \cite{Kachru:2003aw} and large-volume (LVS) \cite{Balasubramanian:2005zx, Conlon:2005ki} approaches \cite{Cicoli:2018kdo, Kachru:2018aqn, Akrami:2018ylq, Moritz:2017xto,Hamada:2018qef,Carta:2019rhx,Hamada:2019ack,Kachru:2019dvo,Demirtas:2019sip,Blumenhagen:2020ire, Demirtas:2020ffz, Blaback:2019ucp}. 

Our focus in this note, however, is on the trans-Planckian issue, and on assessing the evidence for it from the point of view of what we know about EFTs. We here examine whether the trans-Planckian scenario (as illustrated in the right-hand panel of Fig.~\ref{fig:levcross}) plausibly arises from UV physics, using string theory as a guide and, if so, whether it poses a real threat to the robustness of inflationary predictions for the CMB (since, if not, motivation for the TCC largely evaporates). We argue that migration of modes from high to low energies (relative to the Planck scale, say) is very likely to occur in string theory, but that this in itself is not remarkable and need not pose any threat to the robustness of later low-energy predictions. Only the EFT applicable around the epoch of horizon exit of CMB modes is relevant.\footnote{It should be noted that EFT methods applied to gravitational fields generically fail at sufficiently late times, due to `secular growth' effects \cite{Akhmedov:2015xwa,Burgess:2018sou}. This means in particular that perturbative predictions can fail in gravitational fields with horizons, even if $E/M$ remains everywhere very small, but this has nothing to do with TPP effects. Secular-growth issues and late-time evolution can often be understood using RG resummation techniques associated with EFTs for open systems \cite{Burgess:2014eoa,Burgess:2015ajz,Kaplanek:2019dqu} (see also \cite{Burgess:2009bs,Agon:2014uxa,Boyanovsky:2015tba,Braaten:2016sja}), which when applied to cosmology include and extend the late-time resummations associated with stochastic inflation \cite{StochInf,Starobinsky:1994bd,Tsamis:2005hd}.}

In doing so we revisit old arguments \cite{Polchinski:1995ta,Kaloper:2002uj,Burgess:2002ub,Burgess:2003jk} about why the TPP scenario does not pose any particular threat to late-time EFT methods. Our arguments however differ from more recent work that shares our conclusions \cite{Kaloper:2018zgi, Dvali:2020cgt}. In particular, the main arguments we relate, do not actually rely on the specifics of gravity or cosmology at all, and instead arise more generally for time-dependent EFTs across all fields of physics.

\section{EFTs and Time-dependent Gravitational Systems}

We start by summarizing a representative TCC bound, and do so for two reasons. First we wish to remind the reader how the bounds are derived. Second, we wish to emphasize that if the TCC proposal is to be taken seriously it should really be applied at much lower energies than the Planck scale, at least to the extent that UV completions behave as do perturbative string vacua.

\subsection{Sample trans-Planckian bound}

Consider an inflationary scenario with scale-factor $a(t)$ for which the Hubble scale, $H(t) := \dot a/a$ during inflation is approximately constant and is denoted $H_\ssI$. Denote by ${\mfK}(t) := k/a(t)$ the physical momentum of the specific mode with comoving wavenumber $k$, and choose the specific co-moving wavenumber $k_\CMB$ corresponding to the mode that leaves the Hubble scale only to re-enter at the present epoch, so that $\mfK_0 := \mfK(t_0) = k_\CMB/a(t_0) = H_0$, where $H_0 = H(t_0)$ is the present-day Hubble parameter. The time when this mode leaves the Hubble scale during inflation is denoted $t_\CMB$, so $\mfK_\CMB := \mfK(t_\CMB) = k_\CMB/a(t_\CMB) = H_\ssI$. The period between $t_\CMB$ and the end of inflation (at $t = t_e$) is the minimal duration of the inflationary epoch, and during this time the scale factor expands by $N_\CMB = \ln[a(t_e)/a(t_\CMB)]$ $e$-foldings, where
\be
  N_\CMB = \int_{t_\CMB}^{t_{{\rm e}}} H_{\ssI}(t')\, \exd t' \,.
\ee

The TCC asserts that inflation should not be allowed to permit modes with wavelengths shorter than the Planck length, $\ell_p = 1/M_p$, to cross the Hubble scale, and enforcing this in the interval between $t_\CMB$ and $t_e$ means in particular that
\begin{equation}
    \ell_{p} \, e^{N_{{\CMB}}} < \frac{1}{H_{\ssI}} \quad \hbox{and so} \quad e^{N_{\CMB}}<\frac{M_{p}}{H_{\ssI}} \,.\label{eq:TCC}
\end{equation}

More information is available because $N_\CMB$ and $H_\ssI$ are not completely independent of one another, though the relation between them depends on some of the details of what intervenes between the inflationary epoch and today. For instance, suppose for simplicity that $H_\ssI$ does not change appreciably between $t_\CMB$ and $t_e$, and that the energy density driving inflation gets immediately converted with perfect efficiency to thermal energy that is distributed amongst $g_\ssR := g_\star(T_\ssR)$ degrees of freedom at a reheat temperature $T = T_\ssR$, with 
\be \label{TRvsHI}
 T_\ssR^2 \sim H_\ssI M_p \,. 
\ee

Suppose further that after this the universe expands adiabatically and remains radiation dominated right until the relatively recent epoch, $t_{\rm eq}$, of radiation-matter equality, after which it remains matter-dominated right up until very close to the present era. Then the evolution of the mode with $k = k_\CMB$ states
\begin{equation}
 \mfK_{0}^{-1} = \frac{a({t_0})}{a(t_\CMB)} \, \mfK_\CMB^{-1} = \left( \frac{a_{e}}{a_\CMB} \right) \left( \frac{a_{\rm eq}}{a_{e}} \right) \left(\frac{a_{0}}{a_{\rm eq}} \right) \, \mfK_\CMB^{-1} \simeq e^{N_\CMB} \left( \frac{g_{\ssR}^{1/3} T_{\ssR}}{g_{\rm eq}^{1/3}T_{0}} \right) \mfK_\CMB^{-1} \,,\label{eq:kinverse}
\end{equation}
where $a_i := a(t_i)$ for $i = e$, CMB, eq and $0$. The approximate equality evaluates $a_{\rm eq}/a_e$ using conservation of entropy for unit co-moving volume, where $S=a^{3}g_{*}(T)T^{3}$ with $g_*(T)$ denoting the numbers of independent degrees of freedom that are in equilibrium within the cosmic thermal radiation bath, and evaluates $a_0/a_{\rm eq} \simeq T_{\rm eq}/T_0$ using the temperature change during matter domination of the CMB photons.

The desired relation between $N_\CMB$ and $H_\ssI$ comes by using in this last expression the defining conditions $\mfK_\CMB = H_\ssI$ and $\mfK_0 = H_0$; and trading $T_\ssR$ for $H_\ssI$ using \pref{TRvsHI}. This gives
\begin{equation}
e^{N_{{\CMB}}} \simeq \left( \frac{g_{\rm eq}^{1/3}T_{0}}{g_{\ssR}^{1/3} T_{\ssR}} \right)   \frac{H_{\ssI}}{H_{0}} \simeq \frac{T_0}{H_0} \sqrt{\frac{H_\ssI}{M_p}} \sim 10^{29} \sqrt{\frac{H_\ssI}{M_p}}  \,,\label{eq:Ncmb}
\end{equation}
which for the purposes of estimates assumes $(g_\ssR/ g_{\rm eq})^{1/3} \sim \cO(1)$ and the present-day values $T_0 \sim 3 \, \hbox{K} \sim 3 \times 10^{-4}$ eV and $H_0 \sim 100$ km s${}^{-1}$Mpc${}^{-1}\sim 2 \times 10^{-33}$ eV. Taking the logarithm gives the approximate relation $N_{\CMB} \simeq 67+\frac{1}{2}\ln({H_{\ssI}}/{M_{p}})$.

There are two general constraints on $H_\ssI$ that provide upper and lower bounds on $N_\CMB$ independent of the TCC. On one hand, the amplitude of primordial fluctuations revealed by CMB data provides the upper bound $H_{\ssI}<10^{-5}M_{p}$. On the other hand, requiring reheating to happen at temperatures above the weak scale similarly gives $T_\ssR >10^{-15}M_{p}$, which implies $H_{\ssI}>10^{-30}M_{p}$. Together these give 
\be \label{Nrange}
32 \lsim N_{\CMB} \lsim 61.
\ee

A stronger upper bound on $H_\ssI$ can be derived, however, if the TCC condition \pref{eq:TCC} is used in \pref{eq:Ncmb}:
\begin{equation}
H_{I}<10^{-20}M_{P} \,,\label{eq:NcmbTCC}
\end{equation}
a bound that also implies -- from \pref{eq:Ncmb} -- that $N_{{\rm cmb}}<46$. Such low values for $H_\ssI$ usually also imply extremely small slow-roll parameters and so (for single-field models) predict undetectable production of primordial gravitational waves. In particular, the measured scalar power spectrum satisfies
\be
  {\cal P}_{\ssR} = \frac{1}{8\pi^{2}\epsilon} \left(\frac{H_{\ssI}}{M_{p}}\right)^{2}\sim 10^{-9} \,,
\ee
and so $H_\ssI/M_p \lsim 10^{-20}$ implies the slow-roll parameter is bounded by $\epsilon \lsim 10^{-33}$.  

We see from these arguments that permitting no trans-Planckian modes to cross the Hubble scale imposes radical constraints on the kinds of possible inflationary models. And the above reasoning only assumes the minimal inflation required to process observable modes. The constraints only get stronger if there is a longer period of inflation that occurs before the epoch $t=t_\CMB$, becoming impossible to satisfy in the eternal-inflation limit where the epoch of inflation extrapolates arbitrarily far into the past. 

Stronger constraints also arise if the cutoff scale past which descending states pass is lower than $M_p$, and we argue below that this is actually very likely to be true if existing string constructions are a guide as to the nature of the UV completion. The dangerous UV cutoff is lower in these constructions because $\Lambda$ defines the scale above which physics is not properly described using only the states present in the low-energy cosmological description. But if string theory is any guide there are usually a great many UV states (from the point of view of 4D cosmology) whose masses are well below the 4D Planck mass. These include both massive string vibrational modes and Kaluza-Klein excitations of extra-dimensional fields, some of whose masses are summarized below.

\subsection{UV Information and Strings}

Are these constraints really required to avoid problems using EFT methods in cosmology?

One question one might have about the above analysis is whether there is evidence that states really can descend past a fixed cutoff in UV complete theories, and what plausible values for their masses might be. We follow the literature and use string theory as the representative UV completion, since it is well-enough developed enough to ask such questions in detail. Several issues naturally arise.

\subsubsection*{Level crossing in String theory}

Since string theory is UV complete it has no need for arbitrary cutoffs, making explicit that the only scales relevant to setting up a low-energy theory are physical ones, describing particle masses and background-field geometries. Within string theory breakdown of EFT methods might possibly arise from situations where both IR and UV states evolve in energy, such as depicted in the left-hand panel of Fig.~\pref{fig:levcross}. Does this actually occur?

There has long been good evidence that it does. Although all string masses are field-dependent, it is not true that they all depend on these fields in the same way and because of this their mass ratios can change. Toroidal compactifications of flat background geometries provide the simplest examples of this. When one direction of space is a circle of radius $R$ the components of momentum in this direction become quantized so that $p_n = 2\pi n/R$ for $n$ an integer. Because it has length, a string can also wind around the compact circle, possibly an integer $m$ number of times.  In the absence of motion in other spatial directions this leads to a mode energy of the form 
\be
   \cM_{n,m,N}^2 = M_\ssN^2 + \left( \frac{2\pi n}{R} \right)^2 + \left( \frac{m R}{\alpha'} \right)^2 \,,
\ee
where $N$ labels the string internal vibration quantum numbers, $1/\alpha'$ gives the string tension, $M_\ssN^2 \propto 1/\alpha'$ describes the contribution of string vibrational modes, while $n$ and $m$ are the integers labelling the momentum and winding modes for the compact direction.  

Two things matter in this expression. First, in an effective low-energy description the radius $R$ is a propagating modulus field, whose value can evolve slowly in space and time -- in particular allowing $R(t)$. Second, the contributions to the energy of vibrational, winding and momentum modes depend differently on $R$, and they can famously\footnote{Famously because the spectrum is symmetric under the simultaneous interchange $n \leftrightarrow m$ together with the simultaneous replacement $R \leftrightarrow \alpha'/R$ \cite{Kikkawa:1984cp, Sakai:1985cs, Sathiapalan:1986zb,Nair:1986zn}.} change which is heavy and which is light, depending on the value of $R$. In particular, if the UV sector is defined by the string scale ({\it i.e.} the smallest nonzero $M_\ssN$) then for $R \gg \sqrt{\alpha'}$ the winding modes lie in the UV sector while momentum modes belong in the low-energy EFT, and vice versa if $R \ll \sqrt{\alpha'}$. For  particular values of string moduli massive states can descend to become massless, sometimes filling out the gauge degrees of freedom for enhanced symmetries at these points.  

Given all of these properties it seems very plausible that UV/IR level crossing can occur, and if the smallest nonzero $M_\ssN$ is chosen to define the UV regime it is independent of $R$ and so remains unchanged even if $R(t)$ varies slowly in time.  EFT methods that assume energies much below the string scale do indeed break down once $R \sim \sqrt{\alpha'}$, but there is no evidence at all that the EFTs that apply in the regimes $R \gg \sqrt{\alpha'}$ or $R \ll \sqrt{\alpha'}$ cannot be trusted because of this.

\begin{figure}
\centerline{
\put(-50,0){\begin{picture}(100,100)
    \thicklines
    \put(10,40){\line(5,1){100}}
    \put(10,75){\line(3,-1){100}}
    \put(10,77){\line(3,-1){100}}
   \multiput(10,53)(4,0){25}{\circle*{1}}
    \put(-25,80){$M_w(t)$}
    \put(-25,53){$M_v$}
    \put(-25,30){$M_m(t)$}
    \put(20,10){Crossing of states}
    \end{picture}}
}
\caption{Sketch of the adiabatic time evolution for the energy, $M_m(t)$, for a momentum state (single solid line); $M_w$, a winding state (solid double line) as well as the lowest vibrational state, $M_v$ (horizontal dotted line) as a radius modulus $R(t)$ evolves from large to small values (with time flowing to the right). \label{fig:TDcross}}
\end{figure}
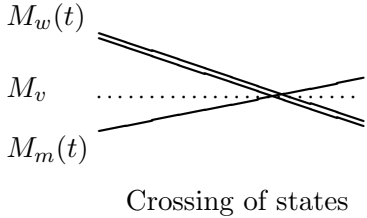

\subsubsection*{Scales in LVS string compactifications}

To drive home the point about $M_p$ usually only being the {\it largest} of many UV scales in string compactifications, we provide here a short summary of the many scales arising in a particular class of string vacua. We use for these purposes the large-volume scenario (LVS) of string compactifications in which a special role is played by the overall compactification volume, ${\cal V}$, written in string units.  Control over the calculation requires this to be large, and in practice typical models have ${\cal V}\gtrsim10^{3}$. In the LVS a large variety of scales arise because the masses of various states depend very differently on $\cV$, generating a cascade of scales in the low-energy theory when $\cV$ is large.

In particular the generic Kaluza-Klein (KK) scale, the string scale and the 4D Planck scale are related by powers of $\cV$ in the following way (keeping track only of powers of $\cV$)
\be
  M_s \sim \frac{M_p}{\cV^{1/2}} \,, \quad
  m_{\KK} \sim \frac{M_{s}}{{\cal V}^{1/6}} \sim \frac{M_{p}}{{\cal V}^{2/3}}
  \quad \hbox{and}\quad m_{3/2} \sim m_{\rm mod} \sim \frac{M_s}{\cV^{1/2}} \sim \frac{M_p}{\cV} \,,
\ee
where $m_{3/2}$ is the gravitino mass and $m_{\rm mod}$ denotes the typical mass of extra-dimensional moduli. For large $\cV$ one is led to the hierachies $m_{3/2} \sim m_{\rm mod} \ll m_{\KK} \ll M_{{s}} \ll M_{p}$. A 4D effective description can only apply at energies well below $m_\KK$ (and therefore also well below $M_p$).

What is attractive about these vacua from the point of view of inflation is the existence of a broad category of moduli -- called `fibre' moduli -- whose mass is generically suppressed relative to all of these because their mass is partially protected by the approximate scaling- and super-symmetries at the lowest order, leading to a suppression of their mass by additional powers of $\cV$, with \cite{Cicoli:2008gp}  $m_{\rm fiber} \sim {M_p}/{\cV^{5/3}}$. This extra suppression allows these moduli to have masses generically of order the Hubble scale, making them natural inflaton candidates.

So far as the TCC is concerned, the EFT describing the low-energy limit of these models should be cut off well below $M_p$: perhaps at $M_s$ for the 10D regime and $m_{\KK}$ for a 4D EFT. Both of these are naturally several orders of magnitude below $M_p$, making the constraints mentioned above even stronger.

\subsection{10D and 4D EFTs}

Most of what is known in practice about string theory itself has been learned using EFT techniques and there seems every reason to expect that whenver one can find a sensible string compactification with a strong hierarchy between the string length, $\ell_s = \sqrt{\alpha'} = 1/M_s$, and other longer scales then the low-energy dynamics will be described by a series of EFTs tailored to each of the relevant scales. Even the recent progress towards the understanding of black hole information loss has been achieved within an EFT of quantum gravity (see for instance \cite{Almheiri:2020cfm} and references therein).

As mentioned above, what complicates this general picture when making explicit 10D constructions is the fact that the stabilization of moduli requires compactifying using complicated geometries, having a multitude of different length-scales, $L_i$, rather than just the single Kaluza-Klein (KK) scale found in the simplest systems. A clean description of the whole geometry completely in terms of an effective 10D field theory is only guaranteed if {\it all} of these length scales satisfy $L_i \gg \ell_s$. When only some are sufficiently large then the 10D field theory description can appear singular in the regions where it breaks down. 

The apparent obstructions to proving the existence of de Sitter solutions in string theory are mostly of this type: it is hard\footnote{For recent progress in this direction see \cite{Andriot:2020vlg,Crino:2020qwk} and references therein.} to find examples with present-day tools for which all length scales in a 10D solution are hierarchically large compared to $\ell_s$. Difficulties finding such purely low-energy solutions are not evidence in themselves that stringy de Sitter solutions do not exist. They ultimately leave this question unanswered: one must compute using the full UV theory to decide. But experience with EFTs in other areas of physics (such as atoms) also tell us that even when one region of a solution requires UV completions for complete understanding, this does not mean all control must be lost over the rest of the solution. 

In particular, all evidence indicates that slow 4D dynamics on scales much longer than those of the compact extra dimensions can be captured by a standard 4D EFT. Time dependence at these low energies can be completely understood within this EFT provided it is adiabatic \cite{Burgess:2003jk,Burgess:2017ytm,TheBook}, which requires in particular that time derivatives of {\it all}\ fields must be small compared with the {\it lowest} extra-dimensional mass scale, 
\be
  |\partial_t \varphi/\varphi| \ll m_\KK := \hbox{min} \; \frac{1}{L_i}
\ee
which includes the usual condition on the 4D Hubble scale: $H \ll m_\KK$. 

Although most would agree that a 4D EFT describes such slow evolution, several features of this 4D EFT are less familiar.

\begin{itemize}
\item It is widely appreciated that the 4D effective theory of gravity comes as an expansion in powers of derivatives of fields, such as $R^n/M^{2n-4}$ (where $R^n$ here schematically denotes all possible scalar contractions involving $n$ powers of the curvature tensor and its derivatives). It is less well known that the scale $M$ with which curvatures are compared is generically {\it not} the 4D Planck mass, $M_p$ for any $n > 2$ \cite{Burgess:2003jk}. As mentioned earlier, perturbative string constructions contain a variety of different scales, with $m_\KK \ll M_s \ll M_p$. But for EFTs the largest UV mass generically only dominates for the relevant (superrenormalizable) interactions, and it is the {\it smallest} UV mass-scale that instead generically dominates for irrelevant (nonrenormalizable) ones, as in the schematic form
\be
  \cL_{\rm eff} \simeq \sqrt{-g} \left[ M_p^2 R + R^2 + \frac{R^3}{M^2} + \cdots \right]\,.
\ee
A similar argument applies for EFTs in higher dimensions, though on dimensional grounds the power of $M$ that applies for a given $n$ differs from the 4D result. In particular one expects $M_s$ to dominate in any 10D EFT (as is indeed found for the $\alpha'$ expansion), and it is $m_\KK$ that should generically dominate the denominators for irrelevant effective couplings in the 4D EFT. (The key word here is `generically', but when it is not the lightest scale that appears in a denominator there is usually a reason for it.)
\item Although the UV scale $M$ controlling the {\it derivative} expansion typically is not given by $M_p$ in 4D EFTs of gravity, the fact that $M_p^2$ appears in front of the leading Einstein-Hilbert action implies that it is true that the {\it loop} (or semiclassical) expansion is controlled by powers of derivatives suppressed by $M_p$ (in 4 dimensions), which for cosmology means loops generically come with\footnote{Although it has been argued that {\it all} corrections are necessarily this size \cite{Kaloper:2002uj}, this need not be strictly true for several reasons. Powers of $H^2/M^2$ can be compared with the lower scales $M \ll M_p$, associated with higher derivative terms (as in bullet 1). There can be systematic suppression by other scales (like $\dot H$), often expressed in terms of dimensionless couplings (like slow-roll parameters). Larger scales than $H$ can arise in the numerator, such as the scale $\mu \gg H$ that must appear in the matter sector that dominates the cosmic energy density to generate a nonzero Hubble scale $H \sim \mu^2/M_p$, since the existence of this scale can allow corrections to arise as powers of $\mu^2/M_p^2 \sim H/M_p$ rather than $H^2/M_p^2$ \cite{Burgess:2002ub}.} powers of $H^2/M_p^2$. 

\item Because the perturbative (and low energy) expansions in string theory are controlled by powers of fields (such as the dilaton), the low-energy EFT for perturbative string theories automatically contain several accidental scaling symmetries \cite{Witten:1985bz,Burgess:1985zz}. The interplay between these scaling symmetries and supersymmetry lies at the root of the ubiquitous appearance of no-scale models in string compactifications \cite{Burgess:2020qsc}, and of the no-go theorems \cite{no-go} that quantify the difficulty finding de Sitter solutions. We believe more systematic exploration of the consequences of these symmetries is likely to be a more effective direction to pursue than are difficult-to-test conjectures. It is encouraging in this regard that the inflationary models towards which they point are among the most successful descriptions \cite{Martin:2013nzq} of the current data (for a more detailed summary of these issues see \cite{Burgess:2016owb}).
\end{itemize}

\section{Time-dependence in EFTs for Non-gravitational Systems}

In this section we describe two field-theory examples that illustrate how many of the time-dependent issues encountered in cosmology also arise in the absence of gravity. These examples also show how nothing in principle precludes EFTs from describing adiabatic motion. 

\subsection{Complex-scalar Model}

To illustrate how EFTs can properly capture time dependent backgrounds, consider the simplest symmetry breaking model: a complex scalar field $\phi=\rho e^{i\theta}$ with the standard Mexican hat potential\footnote{Here we follow the discussion in \cite{TheBook}, see also \cite{Achucarro:2010jv,Achucarro:2010da,Achucarro:2012sm,Shiu:2011qw,Burgess:2012dz}. } 
\be
{\mathcal L}=-\partial^\mu\phi \,\partial_\mu\phi^* - V(\phi)
 \quad \hbox{with} \quad
 V(\phi)=\frac{\lambda}{4}\left(|\phi|^2-v^2\right)^2 \,.
\ee
for which  the symmetry $\theta\rightarrow \theta +$constant implies the conservation of the Noether current:
\be
J_\mu=i\left(\phi\,  \partial_\mu\phi^*-\phi^*\partial_\mu\phi\right)= 2\rho^2\partial_\mu\theta
\ee

Rather than studying small fluctuations around the standard time-independent vacuum, $\rho=v$, giving rise to the fields $\rho$ acquiring a mass $m^2=\lambda v^2$ and massless Goldstone boson $\theta$, we instead consider a time-dependent background of the form:
\be
\rho=\rho_0, \qquad \theta = \theta_0+\omega t
\ee 
with $\rho_0, \theta_0,\omega$ constants that are constrained by the field equations to satisfy:
\be
\rho_0^2=v^2+\frac{2\omega^2}{\lambda}.
\ee
This shows that the time dependence of $\theta$ forces $\rho$ to shift from its vacuum value $\rho=v$ at the minimum of the potential. The Noether current and energy density can be easily computed in this background to give:
\be
J^\mu=-2\omega\left(v^2+\frac{2\omega^2}{\lambda}\right)\, \delta_0^\mu\ , 
\qquad {\mathcal E}=\omega^2\left( v^2+\frac{3\omega^2}{\lambda}\right)\label{jeuv}
\ee

One might ask how the effective theory involving only the Goldstone mode can capture the slowly rolling solution, given that the effective theory has no massive scalar $\rho$ and no scalar potential up which to climb to provide the centripetal acceleration.  To find out, write the EFT at scales below $m$ in terms of the canonically normalised Goldstone boson field $\xi=\sqrt{2}v\theta$ with an effective action:
\be
S_{\rm eff}= -\int d^4x\left[\frac{1}{2}\partial^\mu\xi\partial_\mu \xi-\frac{\lambda}{4m^4}\left(\partial^\mu\xi\partial_\mu \xi\right)^2+\cdots\right]
\ee
where the coefficient of the $(\partial \xi \partial \xi)^2$ term, as well as any higher-order interactions, is determined in the usual way (such as by matching the effective action with the scattering amplitudes computed from the original model expanded about the time-independent vacuum). 

Writing this effective action in terms of $\theta$ (in order to compare with the results above) we compute the  Noether current corresponding to the shift symmetry of $\theta$, leading to
\be
J^\mu_{\rm eff}= 2v^2\partial^\mu\theta \left[1-\frac{2}{m^2}\left(\partial^\mu\theta \partial_\mu \theta \right)+\cdots\right]
\ee
The effective Hamiltonian is similarly computed from ${\mathcal H}_{\rm eff}=\pi_{\rm eff}\, \dot\theta-{\mathcal L}_{\rm eff}$ with $\pi_{\rm eff}={\delta S_{\rm eff}}/{\delta\dot\theta}$ :
\be
{\mathcal H}_{\rm eff}=v^2\dot\theta^2+v^2\nabla\theta\cdot\nabla\theta + \frac{3\lambda v^4}{m^4}\dot\theta^4+\cdots
\ee

This effective theory can also be used to study the energetics of, and the fluctuations about, a slowly rolling solution, $\theta=\theta_0+\omega t$, which is easily seen to be a solution of the equations of motion derived from $S_{\rm eff}$. For this background the Noether current and energy density evaluate to:
\be
J^\mu_{\rm eff}=-2\omega\left(v^2+\frac{2\omega^2}{\lambda}+\cdots\right), \qquad {\mathcal E}= \omega^2\left(v^2+\frac{3\omega^2}{\lambda} + \cdots\right)
\ee
where we use $m^2=\lambda v^2$. Note that these results perfectly match those in equation (\ref{jeuv}), at least out to order $\omega^2$. The first sub-leading term of the EFT proportional to $\left(\partial^\mu\xi\partial_\mu \xi\right)^2$ is precisely what is required to capture the physics of the time dependence of  the UV theory. Higher derivative terms can be included to capture higher-order terms in powers of $\omega/m$, and so this ratio must be small (capturing the solution in the adiabatic regime). This example illustrates how EFTs can robustly capture the low-energy limit of UV physics, even for time-dependent backgrounds. A similar expansion also provides an efficient way to study long-wavelength fluctuations around the corresponding background (as is well-known in the cosmology community, see {\it e.g.}~\cite{Cheung:2007st,Achucarro:2010jv,Achucarro:2010da,Achucarro:2012sm}). 

\subsection{UV descent and Landau Levels}

TCC constraints are ultimately motivated by the requirement that no trans-Planckian modes be allowed to pass through Hubble exit during any inflationary regime. This is motivated, as least implicitly, by the belief that having modes pass through a UV cutoff (as in the right panel of Fig.~\ref{fig:levcross}) would be a bad thing; sufficiently so to be worth taking drastic measures to avoid. 

At first glance this might seem reasonable, and to be required if one is to save the principle of decoupling. After all, if unknown physics can descend to scales relevant for observations, needn't uncontrolled theoretical errors necessarily pollute physical predictions produced using EFTs for the observed scales? Furthermore, if this breakdown should be specific to cosmology and the very early inflationary universe, having decoupling break down in this way might not be inconsistent with evidence for the robustness of decoupling based on experience on Earth in less exotic circumstances.

This section argues that both parts of this reasoning is mistaken, following arguments made long ago \cite{Polchinski:1995ta,Burgess:2003jk}. First, we argue that the issue of modes descending below a fixed cutoff is not at all specific to gravity or to cosmology and generically occurs for EFTs that involve time-dependent background fields. Experience with what happens in non-gravitational contexts helps us understand why merely having modes descend past the UV cutoff does not generically undermine decoupling. That is to say, although examples can be constructed for which UV modes descending past a cutoff {\it can} cause dramatic effects at later times, they only do so when less robust assumptions are made about the nature of the UV physics that is involved.

To establish that gravity is actually peripheral to the central issue, consider the example of a charged particle (with charge $e$, mass $m$) in a constant magnetic field $B$. The energy levels for such a particle are given (in a state with zero momentum parallel to the field) by Landau levels,
\begin{equation} \label{eq:landau}
E_{n}  =\omega_{c}\left( n + \frac12\right),
\end{equation}
where $\,n=0,1,2,\ldots$ and the cyclotron frequency is given by
\begin{equation}
\omega_{c}  =\frac{eB}{m}.\label{eq:cyclotron}
\end{equation}
For any given UV-cutoff scale $\Lambda$ it is clear that these satisfy $E_n > \Lambda$ whenever $n > N \sim \Lambda/\omega_c \gg 1$. 

Issues similar to those that arise in cosmology also occur here if we imagine slowly turning the magnetic field off, so $B = B(t)$ with $\dot B < 0$. We assume the time variation is slow enough to be adiabatic; that is, the magnetic field's Fourier transform $\tilde B(\omega)$ has support only for frequencies\footnote{This also ensures that $\omega \ll \Lambda$ and so this time evolution can be described purely within the low-energy theory  (see {\it e.g.} \cite{Burgess:2003jk,Burgess:2014lwa,Burgess:2017ytm,TheBook} for discussions of the conceptual issues associated with the validity of EFTs in time-dependent situations).} $\omega \ll \omega_c$. In particular this implies
\begin{equation} \label{adiab}
|\dot B| \ll \omega_c B = \frac{eB^2}{m} \,. 
\end{equation}
Among other things, the condition $\omega \ll \omega_c$ ensures that the time-varying magnetic field does not cause transitions between Landau levels with different values for $n$.  In this type of adiabatic regime eq.~(\ref{eq:landau}) continues to hold -- using the instantaneous field $B(t)$ -- as does the picture of Landau levels. 

As $B$ shrinks the process illustrated in the right-hand panel of Fig.~\pref{fig:levcross} takes place, with successive Landau levels descending in energy and a constant stream of nominally UV states passing below $\Lambda$ as time evolves. After sufficient passage of time one can imagine that the magnetic field eventually vanishes: $B \to 0$. In this limit the discrete oscillator states of (\ref{eq:landau}) go over to the continuum of momentum states for a particle in the absence of a magnetic field, as the spacing between the Landau levels collapses to zero. Because all discrete states with fixed $n$ tend to zero energy as $B \to 0$, any continuum state that emerges at vanishing $B$ with nonzero energy $E$ necessarily descends from states whose Landau-level energy was greater than $\Lambda$ in the remote past. 

Naively, this seems to lead to a trans-Planckian problem just as described in the TPP for inflation, but this time one that takes place daily in the lab whenever somebody turns off a magnetic field. To quantify this, suppose at some initial time $t=t_0$ we take $B(t_0) = B_0$.  Let us consider a specific mode number $1\ll n_0 \ll N$ at this initial time, for which
\begin{equation}
E_{n_0}(t_0) \simeq n_0\, \omega_{c}(t_0) =\frac{eB_{0} n_0}{m} \ll \Lambda.\label{eq:cutooff}
\end{equation}
Although this mode is clearly within the low-energy theory at $t=t_0$, at some time $t_{-}< t_0$ this state would be above the cutoff, and so at face value would have been beyond the domain of validity of the EFT that applies only at energies below $\Lambda$:
\begin{equation}
E_{n_0}( t_{-}) = \frac{eB(t_{-}) n_0}{m}  \sim\Lambda \,.\label{eq:Emax}
\end{equation}

For simplicity suppose $\dot{B}$ is constant and so $B(t) = B_0 - |\dot B| t$. With this choice the adiabatic condition $|\dot B| \ll eB^2/m$ eventually fails at very late times as $B$ continues to shrink. The time $t_{\rm nonad}$ when this happens can be pushed into the far future by making $|\dot B|$ smaller and smaller. Requiring $t_{\rm nonad} > t_0$ implies 
\be \label{BdotBound}
  |\dot B| \ll \frac{eB_0^2}{m} \,. 
\ee
When $\dot B$ is constant the time $t_-$ when $E_{n_0}(t)$ reaches $\Lambda$ can be computed and is
\begin{equation}
t_0 - t_{-} =\left(\frac{m\Lambda}{n_0e} - B_{0}\right)\frac{1}{|\dot{B}|} 
\gg \left(\frac{m\Lambda}{n_0e} - B_{0}\right) \frac{m}{e|B_0|^{2}} \,,\label{eq:t-}
\end{equation}
where the final inequality uses \pref{BdotBound}.

What would be the analog of TCC for this system? One would demand that the time
varying field $B_{1}(t)$ must be shut off at the time $t_{-}$ defined by (\ref{eq:Emax}). However this is clearly an artificial requirement, and this simple example suggests why it is not really required (either here or in cosmology).

It is true in this model that states that arrive at low energy with nonzero (but small) energy in the regime $B \to 0$ necessarily must descend from states that initially were at arbitrarily large values of $n$ at earlier times. And at face value the energy of these initial states is higher the further back in time one extrapolates. But we do not need to know the detailed form of the physics at these high energies to conclude robustly that there need not be a late-time disaster as such modes descend from beyond the cutoff. All that is needed to exclude such a disaster is that these states all enter the low-energy regime in their adiabatic vacuum. It does not matter really what this vacuum is, so long as it evolves continuously and adiabatically into the corresponding ground state in the low-energy effective theory.

That is not to say that one cannot arrange for states coming through the cutoff to ruin the low-energy theory at late times. But doing so requires preparing things so that either the adiabatic approximation fails at these earlier times, or that some modes arrive below the cutoff in metastable states rather than being in their adiabatic vacuum. (This latter case might be called a UV `bomb' inasmuch as it is a system that is carefully prepared in a metastable state that stores much energy, which can be released if the adiabatic assumptions that led to its metastability should eventually be violated.) Although such states can be prepared, they are not generic because systems often seek their adiabatic vacuum when perturbed. When they do, no dramatic means are usually required to keep late-time low-energy predictions divorced from early ultraviolet details.

\section{Conclusions}
 
In this note we revisit the validity of EFT in time dependent backgrounds, such as are encountered for  applications to cosmology. In particular we use simple non-gravitational examples to illustrate how many of the issues encountered in cosmology (like descent of UV energy levels and the treatment of intrinsically time-dependent evolution) already arise for much simpler and easier-to-understand systems. 

We use these examples to question the need for cosmological  trans-Planckian conjectures, whose design ultimately aims to avoid supposed problems associated with using EFTs in time-dependent contexts. We argue that EFTs in cosmological backgrounds can be robust at horizon exit even if there were an earlier period during which trans-Planckian modes get stretched out to macroscopic scales. This causes no problems at late times so long as these modes arrive in their adiabatic vacuum.    

\section*{Acknowledgements}
We thank Sebasti\'an C\'espedes, Michele Cicoli, Paolo Creminelli, Francesco Muia, Nemanja Kaloper and Roberto Valandro for useful conversations. CB's research was partially supported by funds from the Natural Sciences and Engineering Research Council (NSERC) of Canada. Research at the Perimeter Institute is supported in part by the Government of Canada through NSERC and by the Province of Ontario through MRI. The work of FQ has been partially supported by STFC consolidated grants ST/P000681/1, ST/T000694/1.

\bibliographystyle{apsrev}

\end{document}